\documentclass[conference]{IEEEtran}
\IEEEoverridecommandlockouts

\usepackage{cite}
\usepackage{amsmath,amssymb,amsfonts}
\usepackage{algorithm}
\usepackage[bookmarks=false]{hyperref}
\usepackage{algpseudocode}
\usepackage{graphicx}
\usepackage{textcomp}
\usepackage{xcolor}
\usepackage{amsthm}
\usepackage{tikz}
\usetikzlibrary{positioning, arrows.meta, shapes.geometric, calc, fit}

\def\BibTeX{{\rm B\kern-.05em{\sc i\kern-.025em b}\kern-.08em
    T\kern-.1667em\lower.7ex\hbox{E}\kern-.125emX}}
\begin{document}

\title{A Capability Broker for Workflow--Network QoS Coordination in B5G/6G Industrial Services
}

\author{

\IEEEauthorblockN{
Qize~Guo\IEEEauthorrefmark{1},
Yan~Chen\IEEEauthorrefmark{1},
Tarik~Taleb\IEEEauthorrefmark{1},
Bjoern~Riemer\IEEEauthorrefmark{2},
Hemant~Zope\IEEEauthorrefmark{2}
and
Hao~Yu\IEEEauthorrefmark{3}
}
\IEEEauthorblockA{\IEEEauthorrefmark{1}Faculty of Electrical Engineering and Information Technology, Ruhr University Bochum, Bochum, 44801, Germany}
\IEEEauthorblockA{\IEEEauthorrefmark{2}Fraunhofer Institute for Open Communication Systems (FOKUS), Berlin, 10589, Germany}
\IEEEauthorblockA{\IEEEauthorrefmark{3}Hangzhou International Innovation Institute, Beihang University, Hangzhou, 31115, China}
\IEEEauthorblockA{Email: \{qize.guo, yan.chen-j3l, tarik.taleb\}@rub.de, \{bjoern.riemer, hemant.zope\}@fokus.fraunhofer.de, haoyu1@buaa.edu.cn}
}
\bibliographystyle{IEEEtran}
\maketitle

\begin{abstract}
Industrial services in beyond-fifth-generation (B5G) and sixth-generation (6G) networks are increasingly executed as multi-phase workflows whose quality-of-service (QoS) demands change across ordered phases. Existing exposure, analytics, and policy-control functions can provide capability information and enforce QoS treatment, but they do not track the admitted QoS agreement between a workflow and the network. This paper studies this missing coordination function in the software control plane. We propose a Capability Broker that represents each admitted trajectory as a lifecycle-managed QoS commitment, including the workflow demand, network capability, validity time, and enforcement state. The Broker operates above existing exposure and policy-control interfaces and enforces capability freshness, duplicate-safe admission, legal state transitions, per-commitment event ordering, recovery routing, and Broker--network state consistency. Prototype results with microbenchmarks, fault injection, ablation, and controlled message loss show that the full Broker preserves commitment consistency with microsecond-level local processing overhead, while removing individual Broker functions exposes duplicate commitments, stale admissions, illegal transitions, lost-update divergences, or Broker--network inconsistency.

\end{abstract}


\section{Introduction}

Industrial services in beyond-fifth-generation (B5G) and sixth-generation (6G) networks increasingly coordinate communication with the execution of an operational workflow, not just with an isolated traffic flow~\cite{intro:5gppp2023InnovationTrends,acia_5g_automation,varga2020iiot5g}. Consider a mobile inspection robot or production-cell controller that first exchanges low-rate telemetry and control commands, then uploads high-rate sensing data or video during inspection, and later switches to reporting or recovery after a defect or network degradation is detected~\cite{intro:XU2025100356,intro:meitz2025literature}. The application may choose among full-quality, reduced-quality, and fallback quality-of-service (QoS) modes for each phase. These choices are coupled: a reduced video mode can extend the inspection phase, lower the confidence of a downstream decision, or constrain which recovery mode remains feasible later. The communication problem is not simply to request one QoS treatment at a time. The control plane also needs to track the admitted QoS agreement as the workflow starts, adapts, recovers, and finishes.

Current 5G systems already provide building blocks. The Network Exposure Function (NEF) exposes selected capabilities and events, the Network Data Analytics Function (NWDAF) provides analytics about service experience and load, and the Policy Control Function (PCF) authorizes QoS treatment, including Alternative QoS Profile (AQP) fallback~\cite{3gpp29522,3gpp23288,3gpp23503,camara_qod}. Their dominant interaction model remains request-driven and session scoped. They can provide capability information and enforce QoS treatment, but they do not define a runtime object that binds a multi-phase workflow demand to a network capability view.

This gap appears when workflow orchestration uses network capability information. Suppose the network provides a capability envelope that describes near-term supportability, and the application submits a demand trajectory that maps upcoming workflow phases to selected QoS modes and fallback options. Once admitted, this trajectory is no longer a single QoS request. It becomes a workflow--network QoS commitment: an admitted agreement that links workflow demand, network capability, validity time, and enforcement state.

Managing this commitment requires a software control-plane function. The Broker must check capability freshness before network-side policy actions, map retransmitted submissions to the same commitment, prevent workflow execution before network-side admission succeeds, route capability changes to recovery, and order asynchronous workflow-side and network-side events under the same commitment identity. Without such an owner, the control plane may use stale capability, create duplicate commitments, accept illegal workflow progress, or lose consistency with the network-side state.

Building on the capability-aware workflow coordination vision in~\cite{capability_commag}, this paper proposes a Capability Broker for workflow--network QoS coordination in industrial B5G/6G services. The novelty is not a new QoS application programming interface (API) or a new workflow scheduler. It is the commitment-management layer required after capability-aware admission has been made. The Broker operates between the Industrial Agent and the Network Agent, turns an admitted demand trajectory into a lifecycle-managed QoS commitment, and keeps this commitment consistent during admission, activation, recovery, release, and abort.

The main contributions are as follows:
\begin{itemize}
\item We formulate workflow--network QoS coordination as a software control-plane problem for capability-aware B5G/6G industrial services. The key abstraction is the capability commitment defined in Section~\ref{sec:context}.

\item We design a Capability Broker that turns exposed network capability from information into a lifecycle-managed service orchestration object. Its six software responsibilities are summarized in Table~\ref{tab:broker_components}.

\item We implement a Broker prototype and evaluate it through local overhead measurement, fault injection, ablation, and lossy signaling experiments.
\end{itemize}

\section{System Context and Commitment Model}
\label{sec:context}

\subsection{Existing Primitives and Missing Object}

5G systems provide analytics, exposure, policy-control, and application-facing QoS APIs for application--network coordination~\cite{3gpp29522,3gpp23288,3gpp23503,camara_qod}. The Service Enabler Architecture Layer for Verticals (SEAL) also defines common enabler services for vertical applications over 3rd Generation Partnership Project (3GPP) networks~\cite{3gpp23434}. These primitives expose metrics, events, QoS profiles, and policy-control entry points. They are necessary for capability-aware industrial services, but they do not define the runtime object created when workflow demand is admitted against network capability.

In this paper, the capability envelope is an abstraction built above existing analytics, exposure, and policy-control functions, not a new 3GPP information element. It may be derived from NWDAF analytics, NEF exposure, CAMARA Quality on Demand (QoD) sessions, operator policy, or local admission estimates. We assume such an envelope is available and focus on the software layer needed after a workflow trajectory is admitted. Without an explicit owner, key lifecycle questions remain open: whether the envelope is fresh, whether a retransmission maps to an existing commitment, whether workflow execution may start, and whether a capability-change event should trigger downgrade, upgrade, recovery, release, or abort.

Softwarized service orchestration systems address a different layer of the problem. Network Function Virtualization Management and Orchestration (NFV MANO) frameworks manage network-service, Virtual Network Function (VNF), Cloud-Native Network Function (CNF), slice, and policy lifecycles. Software-Defined Networking (SDN), Multi-access Edge Computing (MEC), cloud-native orchestration, intent-based networking, and zero-touch management steer traffic, place service components, and automate network-service configuration~\cite{etsi_zsm0091}. This paper studies a narrower object: the QoS commitment created when one workflow demand trajectory is admitted against a network capability envelope.

The Broker also uses known distributed-systems patterns, including idempotent message handling, long-running sagas, durable state, and state repair~\cite{helland2012idempotence,garciamolina1987sagas,terry1995bayou}. These patterns support retried requests, long-running workflows, and desired-state repair. They do not define a workflow--network QoS commitment that links a multi-phase application trajectory to network capability information and policy-control actions.

\subsection{Capability Commitment}

We consider an industrial workflow composed of ordered phases. Each phase may support multiple QoS modes, such as full, reduced, and fallback modes. A demand trajectory specifies how upcoming phases are mapped to selected QoS modes and fallback options. This trajectory may be generated by an application-side planner, a service orchestrator, or an optimization module.

The network side owns capability estimation, policy constraints, and QoS enforcement. It may provide a capability envelope that describes which QoS profiles are expected to be supportable within a planning window. The envelope is not a permanent guarantee or a standardized reservation. It is valid only within its disclosed range and may change when network conditions or admitted demands change.

Once a demand trajectory is admitted against a capability envelope, the Broker represents the result as a capability commitment. A commitment contains a stable commitment identifier, workflow identifier, referenced envelope, idempotency key, validity time, selected trajectory, lifecycle state, and trace identifier. It is narrower than a full workflow model because it records only the communication-related agreement that the Broker must validate and enforce during execution.

The commitment object separates ownership across the three parties. The Industrial Agent (IA) owns workflow semantics, including phase order, mode alternatives, quality tradeoffs, and fallback tolerance. The Network Agent (NA) owns capability estimation and QoS enforcement. The Broker owns the admitted agreement between them.

\subsection{Control-Plane Failures}

Without a lifecycle-managed commitment object, capability-aware workflow orchestration exposes four control-plane failures: stale-envelope admission based on an expired capability view; duplicate commitments caused by retransmitted submissions; illegal transitions such as starting before admission or acknowledging recovery outside the recovery state; and Broker--Network Agent inconsistency after an admission, switch, or release message is lost.

These failures are not data-plane failures. They are software control-plane consistency failures caused by the absence of an explicit owner for the workflow--network agreement. The Broker addresses this gap through four design objectives: commitment ownership, validity-aware policy actions, lifecycle correctness under asynchronous events, and controlled recovery with observability.

We use commitment consistency to mean that an admitted trajectory is fresh, unique under retransmission, lifecycle-legal, consistently ordered under concurrent events, and synchronized with the network-side reservation state.

\section{Broker Design and Lifecycle Semantics}
\label{sec:broker}

\begin{figure}[!t]
  \centering
  \includegraphics[width=\columnwidth]{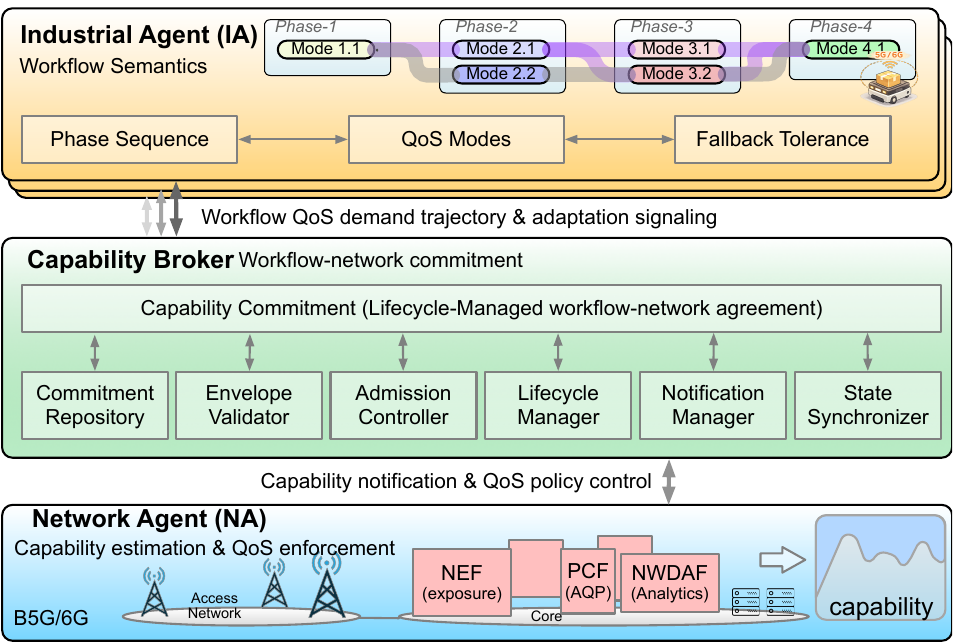}
  \caption{Broker architecture and control paths.}
  \label{fig:framework}
\end{figure}

The Broker is a software control-plane component placed between the IA and the NA, as shown in Fig.~\ref{fig:framework}. It does not replace workflow planning, network analytics, or QoS policy control. It owns the capability commitment defined in Section~\ref{sec:context} and keeps the admitted workflow--network QoS agreement consistent across admission, activation, recovery, release, and abort.

\subsection{Architecture}

The architecture follows the ownership split defined in Section~\ref{sec:context}: the IA owns workflow semantics, the NA owns capability estimation and QoS enforcement, and the Broker owns the admitted agreement between them.

The Broker is intended as an application-function or service-orchestration component behind NEF/CAMARA exposure, not as a replacement for standardized 5G core functions. The NA is an adapter for capability estimation and policy-control calls, keeping the Broker outside the PCF/NWDAF/NEF implementation boundary.

In normal operation, the NA publishes or updates a capability envelope. The IA submits a demand trajectory that maps upcoming workflow phases to selected QoS modes and fallback options. The Broker validates the request, invokes network-side admission through the NA, and creates a capability commitment if admission succeeds. Later workflow progress, capability changes, recovery decisions, and release operations are processed through the same commitment object.

This placement keeps the Broker focused. It does not need to know proprietary workflow logic or compute radio-resource allocation. It manages only the control-plane state required to keep the admitted workflow--network QoS agreement consistent across IA events, NA events, and policy actions.

\subsection{Components and Control Semantics}

The Broker consists of six logical components. The first five enforce local per-event checks on the synchronous IA--Broker--NA path. The State Synchronizer runs outside this path as a background state-checking and repair module.

\begin{table}[!t]
\centering
\caption{Broker software responsibilities.}
\label{tab:broker_components}
\footnotesize
\setlength{\tabcolsep}{2.5pt}
\begin{tabular}{p{0.28\linewidth} p{0.32\linewidth} p{0.27\linewidth}}
\hline
\textbf{Component} & \textbf{Software responsibility} & \textbf{Prevented failure} \\
\hline
Commitment Repository & Durable commitment state & Untracked agreement \\
Envelope Validator & Freshness before policy action & Stale admission or recovery \\
Admission Controller & Duplicate-safe admission & Duplicate commitment \\
Lifecycle Manager & Table-driven state transition & Illegal workflow progress \\
Notification Manager & Event ordering per commitment & Lost-update race \\
State Synchronizer & State checking and repair & Broker--NA inconsistency \\
\hline
\end{tabular}
\end{table}

The Commitment Repository stores the durable state of each pending or admitted commitment. The Envelope Validator checks whether the referenced capability view is still valid before the Broker invokes any NA-side policy action. The Admission Controller maps retransmitted submissions with the same idempotency key to the existing commitment. The Lifecycle Manager applies a transition table to prevent workflow progress outside the admitted state sequence. The Notification Manager serializes events per commitment so that concurrent workflow-side and network-side events cannot overwrite each other's state updates.

The State Synchronizer handles failures that local checks cannot observe. It compares the Commitment Repository with the NA reservation set and repairs disagreements. In production, NA-only reservations should be checked, retried, or reported before any active QoS flow is removed. Deterministic force-release is used only in the prototype to measure convergence. The local components preserve Broker-side consistency, while the State Synchronizer bounds Broker--NA inconsistency caused by lost NA messages. Periodic repair is sufficient because the Broker owns commitment state and the NA exposes applied reservation state.

The Broker exposes two groups of interfaces. The IA-facing interface receives demand trajectories, workflow start events, phase progress events, adaptation choices, completion events, and abort requests. The NA-facing interface receives capability envelopes, admission results, capability-change notifications, policy activation results, profile-switch results, and release acknowledgements. A demand trajectory first creates a pending commitment candidate. Workflow progress is accepted only if it matches the current commitment state. A capability-change event is routed to recovery instead of directly overwriting the admitted trajectory.

\subsection{Commitment Lifecycle}

\begin{figure}[!t]
  \centering
  \includegraphics[width=\columnwidth]{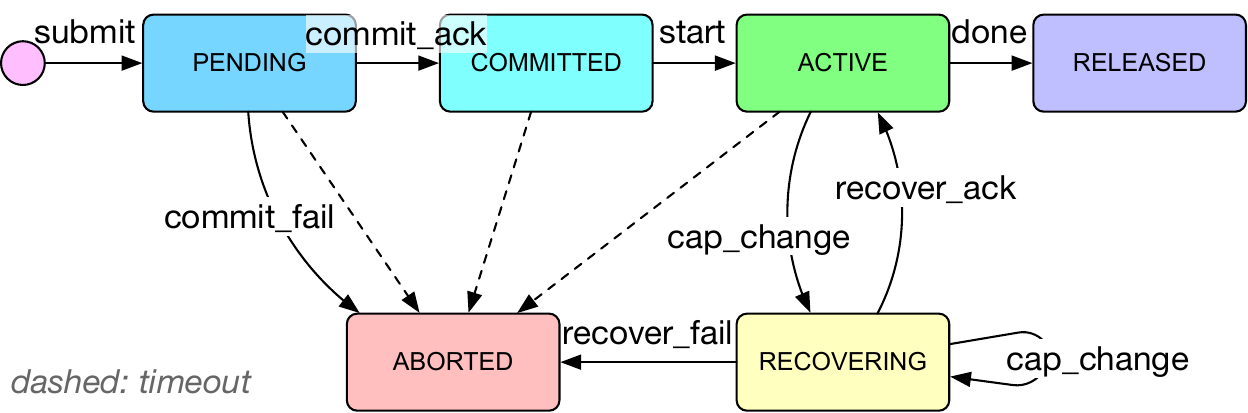}
  \caption{Commitment lifecycle.}
  \label{fig:transtate}
\end{figure}

Fig.~\ref{fig:transtate} shows the lifecycle of a capability commitment. The Broker first creates a commitment in the \texttt{Pending} state when the IA submits a demand trajectory. It then validates the referenced capability envelope, checks the idempotency key, and requests network-side admission through the NA. If admission succeeds, \texttt{commit\_ack} moves the commitment to \texttt{Committed}; otherwise, \texttt{commit\_fail} moves it to \texttt{Aborted}.

The \texttt{Committed} state means that the trajectory has been admitted and the corresponding network-side QoS support has been prepared. A \texttt{start} event moves the commitment to \texttt{Active}. In this state, the workflow executes under the admitted trajectory, and phase progress is checked against the commitment. A \texttt{done} event moves the commitment to \texttt{Released}, where associated QoS policy state is released.

A runtime capability change may affect an admitted segment. Then, \texttt{cap\_change} moves the commitment from \texttt{Active} to \texttt{Recovering}. The Broker requests an adaptation decision from the IA, validates it against the latest capability view, and triggers the corresponding QoS policy update through the NA. If recovery succeeds, \texttt{recover\_ack} returns the commitment to \texttt{Active}; otherwise, \texttt{recover\_fail} moves it to \texttt{Aborted}. A \texttt{timeout} event may also abort any non-terminal commitment. Both \texttt{Released} and \texttt{Aborted} are terminal states; later events are ignored safely and logged.

\subsection{Protocol Invariants}

The first five components implement a uniform acceptance check. For an event $e$ associated with commitment $c$ in current state $s_c$, the Broker accepts $e$ only if
\begin{equation}
\label{eq:accept}
\begin{aligned}
\textsc{Accept}(e, c) ={}&
(\textsc{IsCapChg}(e) \vee \textsc{Fresh}(e.\textit{env}, e.\textit{seg}))\\
&\wedge(\neg\textsc{IsAdm}(e) \vee \textsc{Dedup}(e.\textit{key}))\\
&\wedge\textsc{Legal}(s_c, e),
\end{aligned}
\end{equation}
where \textsc{IsCapChg} identifies capability-change notifications, \textsc{IsAdm} identifies admission requests, \textsc{Fresh} is the Envelope Validator's freshness check on the referenced segment, \textsc{Dedup} is the Admission Controller's idempotency-key lookup, and \textsc{Legal} is the Lifecycle Manager's transition-table check. If \textsc{Accept}$(e,c)$ is false, the Broker rejects the event before issuing any NA-side policy action. The Notification Manager guarantees that this evaluation is atomic per commitment, so two concurrent events on the same $c$ cannot both observe the same $s_c$ and write incompatible post-states.

Three invariants follow. \textbf{I1: No policy action from stale capability.} No admission, recovery, or profile-switch action reaches the NA unless \textsc{Fresh} holds for the affected segment at the moment of action. The \texttt{cap\_change} event is exempt from this check because it signals that the envelope has changed; freshness is checked again at \texttt{recover\_ack}, after the IA submits an adaptation that fits the new envelope.

\textbf{I2: One commitment per idempotency key.} For a given idempotency key, \textsc{Dedup} returns the existing commitment instead of creating a new one~\cite{helland2012idempotence}. A retransmission whose trajectory fingerprint differs from the stored one is rejected as a key collision.

\textbf{I3: Legal state progression.} A commitment state can change only through an event that is legal under the lifecycle transition table. In particular, a capability-change event first routes the commitment to \texttt{Recovering} instead of directly changing the admitted trajectory.

The acceptance predicate is local. It cannot detect that an earlier NA-side action was lost. The State Synchronizer complements it with a state repair loop~\cite{terry1995bayou}, which is evaluated in Section~\ref{sec:evaluation}.

\section{Prototype and Evaluation}
\label{sec:evaluation}

\subsection{Prototype and Workloads}

The evaluation focuses on control-plane consistency rather than radio-resource efficiency. The Broker is designed to keep an admitted workflow--network QoS agreement consistent under stale capability, retransmission, asynchronous events, and message loss toward the Network Agent.

We implement the Broker prototype in Python with an IA driver, an NA surface, a Commitment Repository, an Envelope Validator, an Admission Controller, a Lifecycle Manager, a Notification Manager, and a State Synchronizer. The IA submits demand trajectories and workflow progress events, while the NA publishes capability envelopes, returns admission decisions, and receives QoS policy calls for admission, activation, profile switching, and release.

We validate the prototype in a hybrid setup. The NA surface is aligned to the Fraunhofer Institute for Open Communication Systems (FOKUS) 5G test platform~\cite{6gsandboxBerlinPlatform}, while the IA is simulated through our digital twin box~\cite{iotm_dtb}. We use real \texttt{pcf\_applied} latency samples collected on that platform to calibrate the order of magnitude of the NA-side policy delay. Ten samples from a four-phase profile schedule show two regimes: warm-cache profile changes complete in 62--214\,ms, while cold-cache profile installation requires 2.18--2.31\,s. These samples are not used as a statistical benchmark of the FOKUS platform. They show that the NA-side policy delay is much larger than the Broker-side processing prefix.

The consistency experiments use deterministic callbacks for repeatability. This setup allows retransmission, expired envelopes, illegal lifecycle events, concurrent workflow/network events, and controlled message loss to be reproduced. A trace replay confirms that the full Broker follows the expected path \texttt{Pending} $\rightarrow$ \texttt{Committed} $\rightarrow$ \texttt{Active} $\rightarrow$ \texttt{Recovering} $\rightarrow$ \texttt{Active} $\rightarrow$ \texttt{Released} under a three-phase workflow with one capability-change event. The same replay verifies that a capability-change event is routed through \texttt{Recovering} instead of directly changing the admitted trajectory.

\subsection{Local Overhead}

We first measure local Broker overhead for admission processing, envelope validation, idempotency lookup, lifecycle transition checking, and event dispatch. Each operation is invoked 10{,}000 times under 1, 10, and 100 active commitments. Latency is measured using a monotonic high-resolution timer.

Across all three load levels, admission processing measures 5.4--5.8\,$\mu$s at the 50th percentile (P50) and 7.5--7.6\,$\mu$s at the 95th percentile (P95). Event dispatch measures 0.88--0.96\,$\mu$s P50. The three pure checks, namely envelope freshness, idempotency lookup, and lifecycle transition, each complete in 0.08--0.17\,$\mu$s P50. None of the five operations grows with the number of active commitments because the internal lookups are hash-indexed.

These local costs are several orders of magnitude smaller than the calibrated NA-side policy delay. The Broker's per-admission prefix of approximately 5.5\,$\mu$s is about $4\times10^{-3}$\% of the warm-cache median and $2.5\times10^{-4}$\% of the cold-cache median. Under the measured policy-delay regime, the Broker does not dominate end-to-end commitment latency.

\subsection{Fault-Injection Test}

The main evaluation tests whether each Broker function prevents its corresponding failure class. This is a fault-injection test, not a claim that the injected failure rates match real deployment rates. We construct a 10{,}000-operation mixed workload with deterministic randomization. The workload contains 5{,}000 normal lifecycle events, 1{,}500 retransmitted submissions sharing an idempotency key, 1{,}500 submissions referencing an expired envelope, 1{,}000 illegal-transition probes from \texttt{Committed}, and 1{,}000 concurrent race operations on active commitments. Each illegal-transition probe attempts three illegal events, and each race operation issues a concurrent \texttt{cap\_change}/\texttt{done} pair.

We compare the full Broker with four ablated variants:
\begin{itemize}
\item \textbf{No Idempotency}: retransmitted demand submissions are not deduplicated.
\item \textbf{No Envelope Validator}: envelope validity is not checked before admission or recovery.
\item \textbf{No Lifecycle Manager}: events are applied without checking the legal transition table.
\item \textbf{No Notification Manager}: concurrent workflow-side and network-side events are not serialized per commitment.
\end{itemize}

We measure four failure classes. A \emph{duplicate commitment} occurs when the same idempotency key creates more than one commitment. A \emph{stale admission} occurs when a trajectory is admitted against an expired envelope. An \emph{illegal transition} occurs when an out-of-table lifecycle transition is accepted. A \emph{state divergence} occurs when concurrent events produce inconsistent workflow-side and network-side interpretations of the commitment state.

\begin{table}[!t]
\centering
\caption{Fault-injection ablation results.}
\label{tab:ablation}
\footnotesize
\setlength{\tabcolsep}{1.5pt}
\begin{tabular}{@{}p{0.33\linewidth} p{0.14\linewidth} p{0.12\linewidth} p{0.13\linewidth} p{0.15\linewidth}@{}}
\hline
\textbf{Variant} & \textbf{Duplicate} & \textbf{Stale} & \textbf{Illegal} & \textbf{Divergence} \\
\hline
No Idempotency & 1500 & 0 & 0 & 0 \\
No Envelope Validator & 0 & 1500 & 0 & 0 \\
No Lifecycle Manager & 0 & 0 & 3000 & 1000 \\
No Notification Manager & 0 & 0 & 0 & 945 \\
Full Broker & 0 & 0 & 0 & 0 \\
\hline
\end{tabular}
\end{table}

\begin{figure}[!t]
  \centering
  \includegraphics[width=\columnwidth]{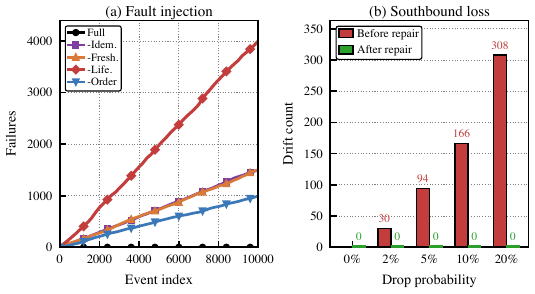}
  \caption{Consistency under injected faults and message loss.}
  \label{fig:consistency}
\end{figure}

Table~\ref{tab:ablation} and Fig.~\ref{fig:consistency}(a) show that each removed function exposes its associated failure class. Without idempotency, retransmissions create duplicate commitments. Without envelope validation, expired-envelope submissions become stale admissions. Without lifecycle validation, each illegal-transition probe accepts three illegal events, giving 3{,}000 illegal transitions; concurrent race operations add 1{,}000 state divergences. Without event ordering, the unsafe variant exposes 945 lost-update divergences, where racing events read the same pre-state and write incompatible post-states.

The full Broker rejects all injected failures and preserves commitment consistency throughout the workload. The divergence counts reflect the intentionally opened race windows in the ablated variants, not real deployment rates.

\subsection{Broker--NA Inconsistency and State Synchronization}

The previous experiments assume that the NA receives all Broker-issued policy calls. In practice, messages toward the NA may be delayed or lost before the Broker observes a successful network-side state change. We measure inconsistency at the end of the workload, after every commitment has been driven through admission, start, and completion. A commitment is inconsistent if the Broker and NA disagree about whether its reservation is still held.

We evaluate this effect with an NA simulator that maintains its own reservation table and drops both request and acknowledgement messages with probability $p$ applied symmetrically to each leg. We sweep $p \in \{0,\,0.02,\,0.05,\,0.10,\,0.20\}$. At each loss level, we replay a 1{,}000-commitment lifecycle workload and measure inconsistency twice: once after the workload, and once after a single State Synchronizer pass.

Fig.~\ref{fig:consistency}(b) shows that without state synchronization, inconsistency grows with the loss rate: 0, 30, 94, 166, and 308 inconsistent commitments out of 1{,}000 at $p=$ 0\%, 2\%, 5\%, 10\%, and 20\%, respectively. At the same loss levels, the effective admission success rate falls from 100\% to 96.8\%, 91.8\%, 82.4\%, and 66.3\% because reservation requests or acknowledgements are silently dropped. State synchronization does not recover lost admission opportunities, but it exposes and repairs residual state disagreement. After one State Synchronizer pass, inconsistency drops to zero at every tested loss level; a second immediate pass detects zero residual inconsistency.

\subsection{State Synchronization Cost and Tunability}
\label{sec:eval_recon_cost}

The previous experiment runs state synchronization only at the end of the workload, whereas a deployed Broker would invoke it periodically. We evaluate both scalability and tunability. For scalability, we time 200 synchronizer invocations for $N \in \{1, 10, 100, 1{,}000, 10{,}000\}$ active commitments, with either no inconsistency or 10\% NA-only inconsistency. For tunability, we replay the 1{,}000-commitment workload at $p=0.10$ and trigger synchronization every $K \in \{10, 30, 100, 300, 1{,}000, 3{,}000\}$ events.

\begin{figure}[!t]
  \centering
  \includegraphics[width=\columnwidth]{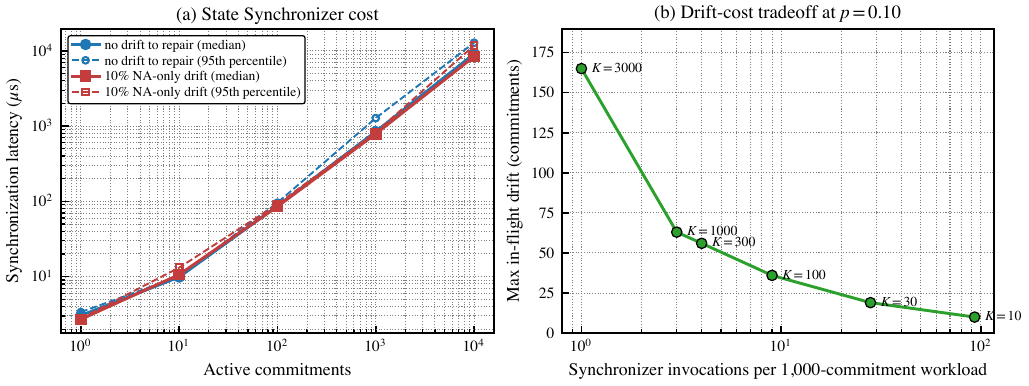}
  \caption{State Synchronizer cost and temporary inconsistency.}
  \label{fig:reconciler}
\end{figure}

Fig.~\ref{fig:reconciler}(a) shows approximately linear scaling: P50 latency rises from 3\,$\mu$s at $N=1$ to about 9\,ms at $N=10{,}000$, with P95 within about $1.5\times$ of P50. Fig.~\ref{fig:reconciler}(b) shows that $K=100$ keeps maximum temporary inconsistency at most 36 commitments with 9 invocations over the workload, smaller $K$ reduces inconsistency further at higher background cost.

Overall, the results show that the Broker's local checks add lightweight processing overhead, while its semantic checks and synchronizer prevent the main control-plane consistency failures introduced by stale capability, retransmission, illegal lifecycle events, concurrent notifications, and lossy signaling toward the NA. The prototype isolates Broker-side consistency from radio-resource scheduling and assumes one Broker instance. Production deployment requires integration with NEF/PCF/NWDAF and replicated Broker state.

\section{Conclusion}
\label{sec:conclusion}

This paper identified workflow--network QoS coordination as a missing commitment-management function in the software control plane of capability-aware industrial B5G/6G services. The proposed Capability Broker represents each admitted trajectory as a lifecycle-managed QoS commitment and enforces freshness, duplicate-safe admission, legal state progression, per-commitment ordering, recovery routing, and Broker--network state synchronization. Prototype results show that removing Broker semantics exposes duplicate commitments, stale admissions, illegal transitions, lost-update divergences, or Broker--NA inconsistency, while the full Broker preserves commitment consistency with microsecond-level local checks and a linearly scalable State Synchronizer. Future work will extend the Broker toward multi-domain coordination, probabilistic capability envelopes, durable Network-Agent signaling, replicated deployment, and joint management of communication and edge-computing resources.

\section*{Acknowledgment}

This work was supported in part by the German Federal Ministry of Research, Technology and Space (BMFTR) through 6GEM+ under Grant~16KIS2411; and by the European Union's Horizon Europe programme through 6G-Path (Grant No.~101139172) and 6GSandbox (Grant No.~101096328). This research was also partially conducted at ICTFICIAL Oy. The paper reflects only the authors’ views, and the European Commission bears no responsibility for any utilization of the information contained herein.

\bibliography{bibtex/IEEEabrv,bibtex/conf}

\end{document}